\documentclass[12pt]{article}
\usepackage{epsfig}
\usepackage{amsmath}
\usepackage{hhline}
\usepackage{amssymb}
\usepackage{times}

\newlength{\dinwidth}
\newlength{\dinmargin}
\begin{document}

\begin{titlepage}

\vspace*{2.0cm}

\begin{center}
\begin{LARGE}

{\bf On the Structure of the Proton, the Photon,}

\vspace*{0.1cm}

{\bf and Colour Singlet Exchange}

\end{LARGE}

\vspace{1.7cm}

{\large Martin Erdmann} \\

\vspace{1.2cm}

\noindent
Institut f\"ur Experimentelle Kernphysik,
Universit\"at Karlsruhe, \\ Engesserstr. 7, 
D-76128 Karlsruhe,
Martin.Erdmann@desy.de \\
\end{center}

\vspace{2cm}

\begin{abstract}
Structure function measurements of the proton, the photon, 
and of colour singlet exchange are analysed using fits
based on logarithmic $Q^2$-dependence.
The fits reveal the $x$-dependencies of the hadronic structures and 
the interaction dynamics as directed by the data.
\end{abstract}

\vspace{2.5cm}


\end{titlepage}

\section{Motivation}
\noindent
A large number of measurements on the structure of different hadronic objects has
been performed in the past decades 
and elaborate descriptions of them have been developed in
terms of parton distribution functions. 
Frequently, the many data points and multiple parameter descriptions make reading of 
physics messages from the data not obvious.
This contribution attempts an
intuitive description of structure functions which has a relation 
to QCD predictions and is applied to structure function
measurements of the proton, the photon, and colour singlet exchange.

\section{Quark Density and Scaling Violations}
\noindent
The QCD evolution equations predict that measurements of hadronic
structures depend on the logarithm of the resolution scale $Q^2$ at which
the structure is probed. 
On this basis, the following ansatz to analyse the $x$-dependence of
structure function data is explored, where $x$ denotes the 
Bjorken fractional momentum of the scattered parton 
relative to the hadronic object:
\begin{equation}
F_2(x,Q^2) = a(x) \; \left[ \ln{\left(\frac{Q^2}{\Lambda^2}\right)} 
\right]^{\;\; \kappa(x)} \; .
\label{eq:f2}
\end{equation}
Here $\Lambda$ is a scale parameter, $a$ reflects 
the charge squared weighted
quark distributions extrapolated to $\ln{(Q^2/\Lambda^2)}=1$, 
and $\kappa$ determines the positive and negative scaling violations of $F_2$.

\begin{figure}[hhh]
\setlength{\unitlength}{1cm}
\begin{picture}(16.0,21.0)
\put(3.6,20.3){\Large a)}
\put(3.6,9.8){\Large b)}
\put(1.0,10.0)
{\epsfig{file=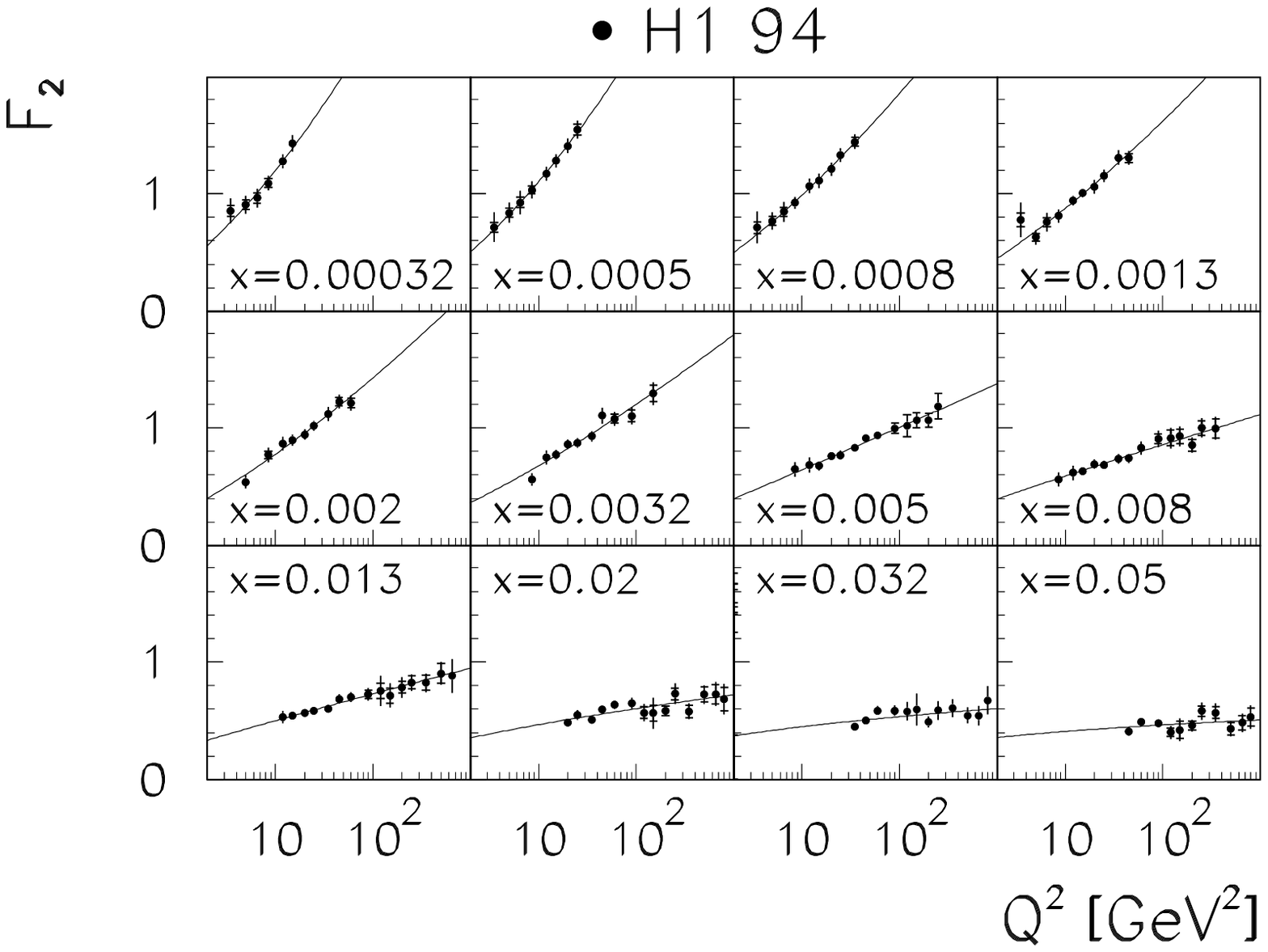, width=15cm}}
\put(1.0,-0.5)
{\epsfig{file=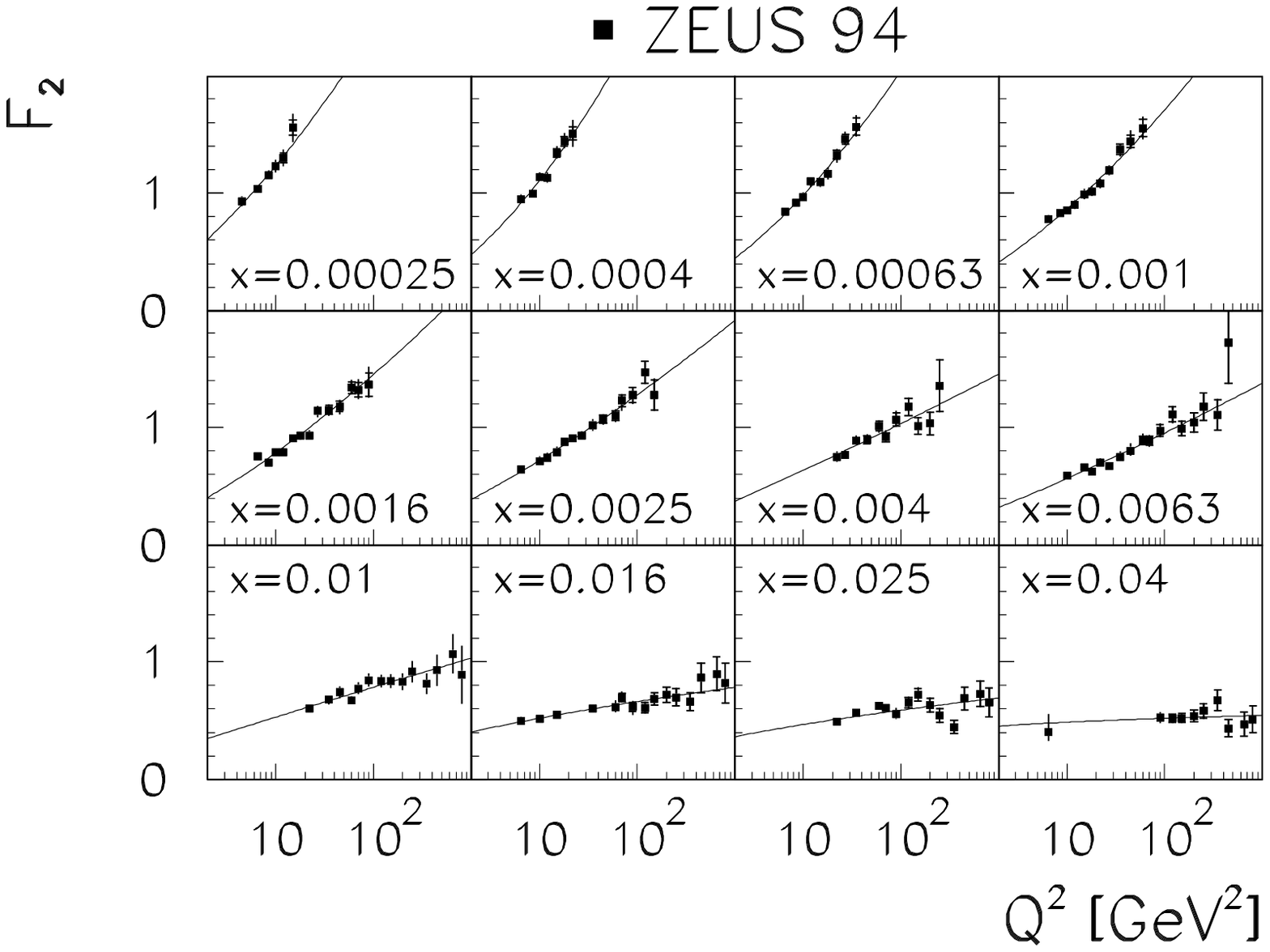, width=15cm}}
\end{picture}
\caption{
\label{fig:proton-hera} 
a) H1 and b) ZEUS 
measurements of the proton structure function $F_2$ are
compared to the two-parameter fits of the normalization term $a$ and 
the scaling violation term $\kappa$ according to eq.~(\protect\ref{eq:f2})
using a fixed value of $\Lambda=0.35$ GeV (full curves).}
\end{figure}

\begin{figure}[hhh]
\setlength{\unitlength}{1cm}
\begin{picture}(16.0,13.0)
\put(1.0,-0.5)
{\epsfig{file=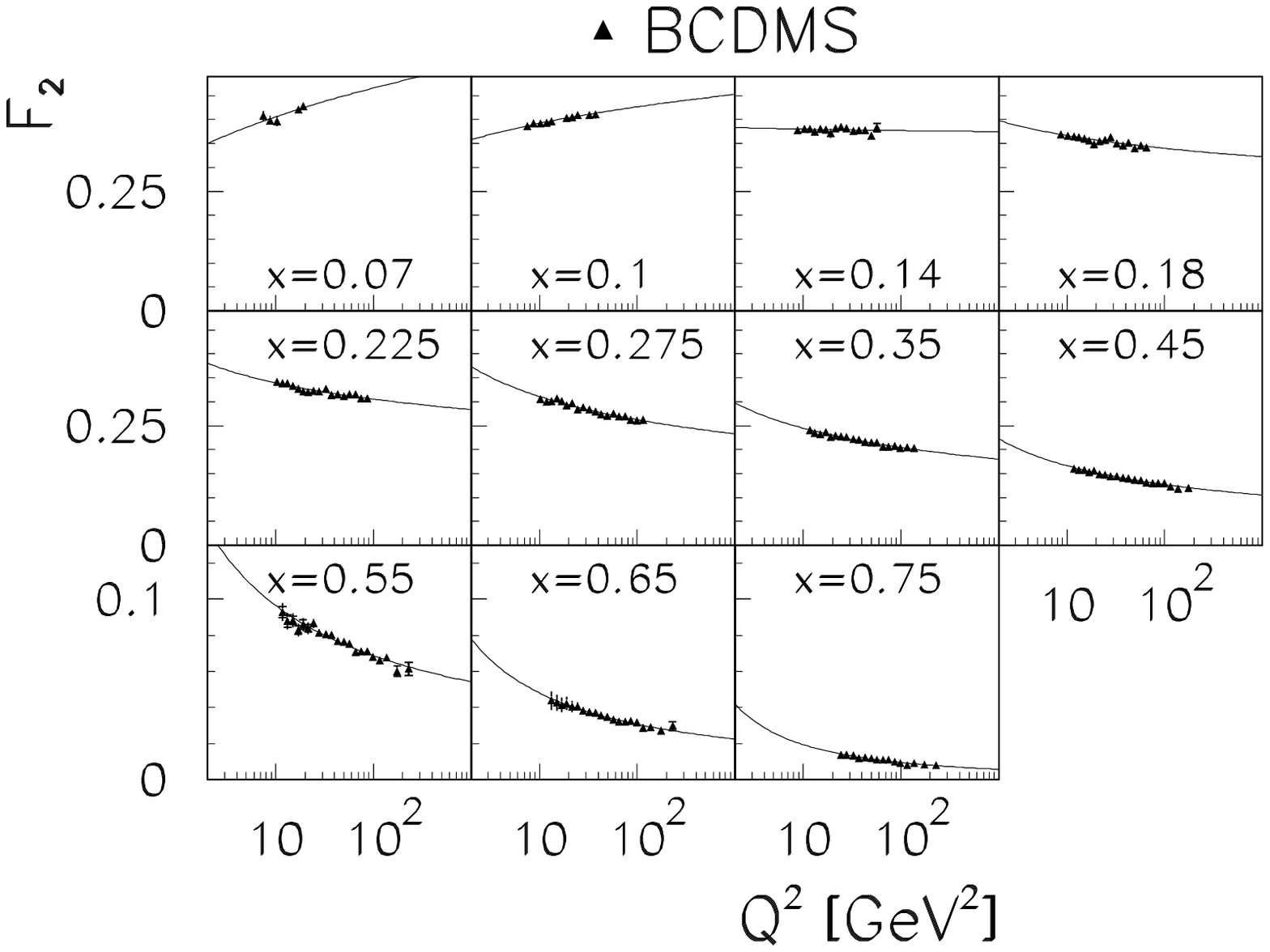, width=15cm}}
\end{picture}
\caption{
\label{fig:proton-bcdms} 
BCDMS measurements of the proton structure function $F_2$ are
compared to the two-parameter fits of the normalization term $a$ and 
the scaling violation term $\kappa$ according to eq.~(\protect\ref{eq:f2})
using a fixed value of $\Lambda=0.35$ GeV (full curves).}
\end{figure}

In Fig.\ref{fig:proton-hera}, H1 \cite{h194} and ZEUS \cite{zeus94} low-$x$
data of the proton structure function $F_2$ for $Q^2>2$\, GeV$^2$ are shown.
In each $x$-bin, the result of a two-parameter fit according to eq.~(\ref{eq:f2})
is shown, using a fixed value of $\Lambda=0.35$ GeV
which represents a typical value of the strong interaction scale.
Only the total experimental errors have been used, ignoring 
correlations between individual data points.
The same fitting procedure has been applied to BCDMS data \cite{bcdms}
which are taken here as a reference sample for the high-$x$ region
(Fig.\ref{fig:proton-bcdms})
(after finalising this analysis the author's attention was called to 
a study fitting the BCDMS data to a similar expression
\cite{mil}).
All data turn out to be well described
($\chi^2/$ndf: $59/107$ H1, $112/115$ ZEUS, $83/155$ BCDMS). 

\begin{figure}[hhh]
\setlength{\unitlength}{1cm}
\begin{picture}(16.0,18.0)
\put(3.8,16.7){\Large a)}
\put(3.8,3.5){\Large b)}
\put(0.5,-0.2)
{\epsfig{file=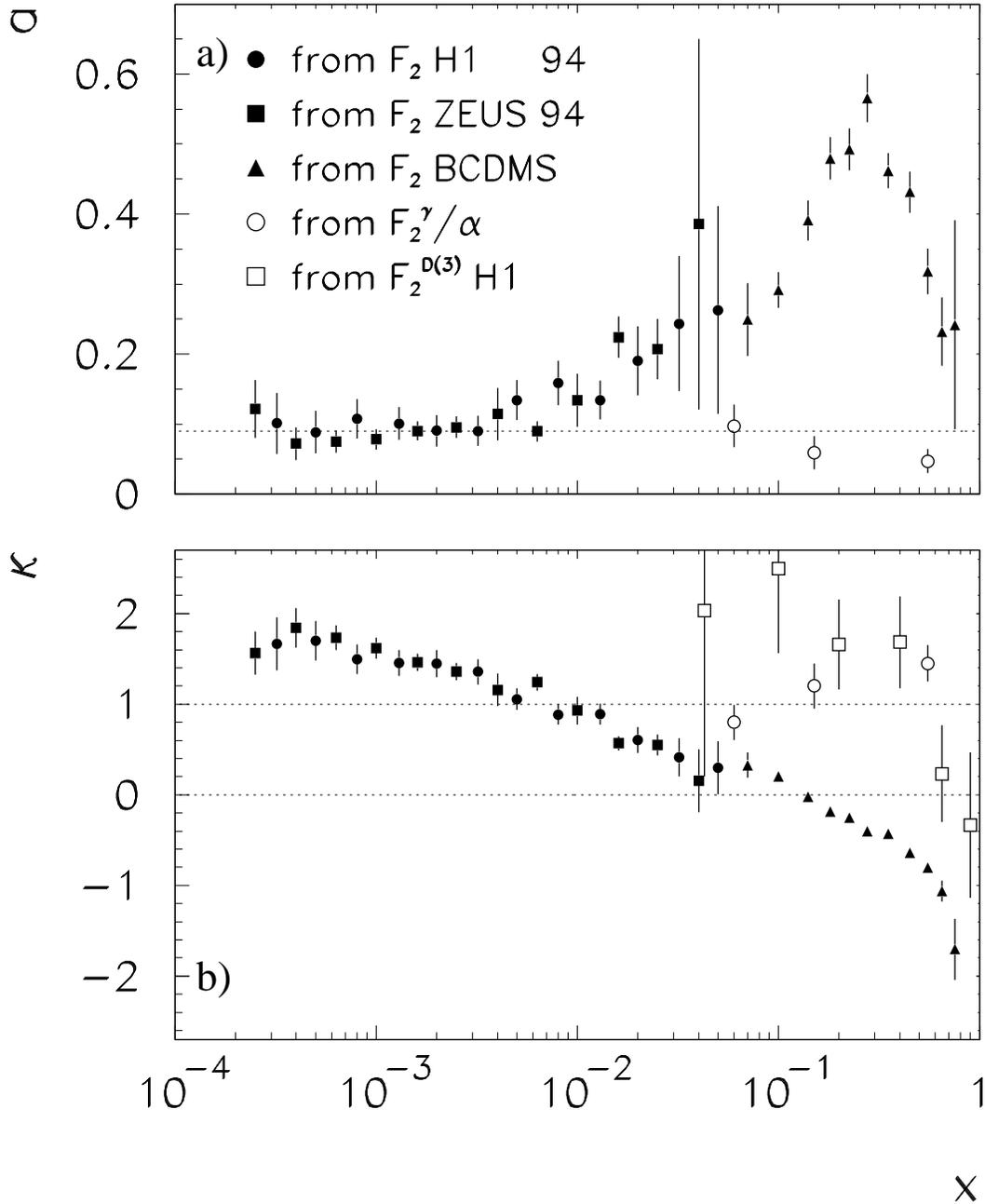, width=16cm}}
\end{picture}
\caption{\label{fig:akappa} 
Summary of the two-parameter fits according to eq.~(\protect\ref{eq:f2}) from 
Figs. \protect\ref{fig:proton-hera}, \protect\ref{fig:proton-bcdms} and 
\protect\ref{fig:fluct}:
a)~the parameter $a$ reflects the valence quarks in the proton
around $x\sim 0.3$ and appears to be constant for the data 
below $x\sim 0.004$ (H1 full circles, ZEUS full square symbols,
BCDMS triangle symbols).
The dotted line serves to guide the eye.
b) The parameter $\kappa$ of the proton shows the 
positive and negative scaling violations below and above $x=0.1$.
Measurements of the photon structure function (open circles) 
exhibit positive scaling violations with $\kappa\sim 1$ 
as expected from QCD.
The open square symbols represent the scaling violations $\kappa$
observed in H1 structure function data of colour singlet exchange
(here $a$ is not shown, see text).}
\end{figure}

The resulting parameters $a$ and $\kappa$ are summarized in 
Fig.\ref{fig:akappa} as a function of $x$. 
For $a$, the data fits exhibit two distinct regions: 
around $x\sim 0.3$ 
they reflect the valence quark distributions.
At low $x$, $a(x)$ is
compatible with converging to a constant value.
The resulting scaling violation term $\kappa$ appears to rise as $x$ 
decreases, exhibiting the negative and positive scaling violations of $F_2$
for $x$ above and below $0.1$ respectively.
The errors in Fig.\ref{fig:akappa} represent the statistical errors of 
the fits.
Both parameters $a$ and $\kappa$ are anti-correlated as can be seen from 
neighbouring points.
No significant $Q^2$-dependence of $a$ and $\kappa$
has been found when the fits were repeated for two intervals in $Q^2$ 
(above and below $20$ GeV$^2$).

With $a$ being approximately constant below $x\sim 0.004$, 
changes of $F_2$ at low $x$ result from the scaling violation 
term $\kappa$ alone, indicative of 
the interaction dynamics that drives $F_2$ and
in support of \cite{lowx,das,grv}.

\begin{figure}[hhh]
\setlength{\unitlength}{1cm}
\begin{picture}(16.0,17.0)
\put(3.4,17.7){\Large a)}
\put(3.4,8.7){\Large b)}
\put(1.0,9.0)
{\epsfig{file=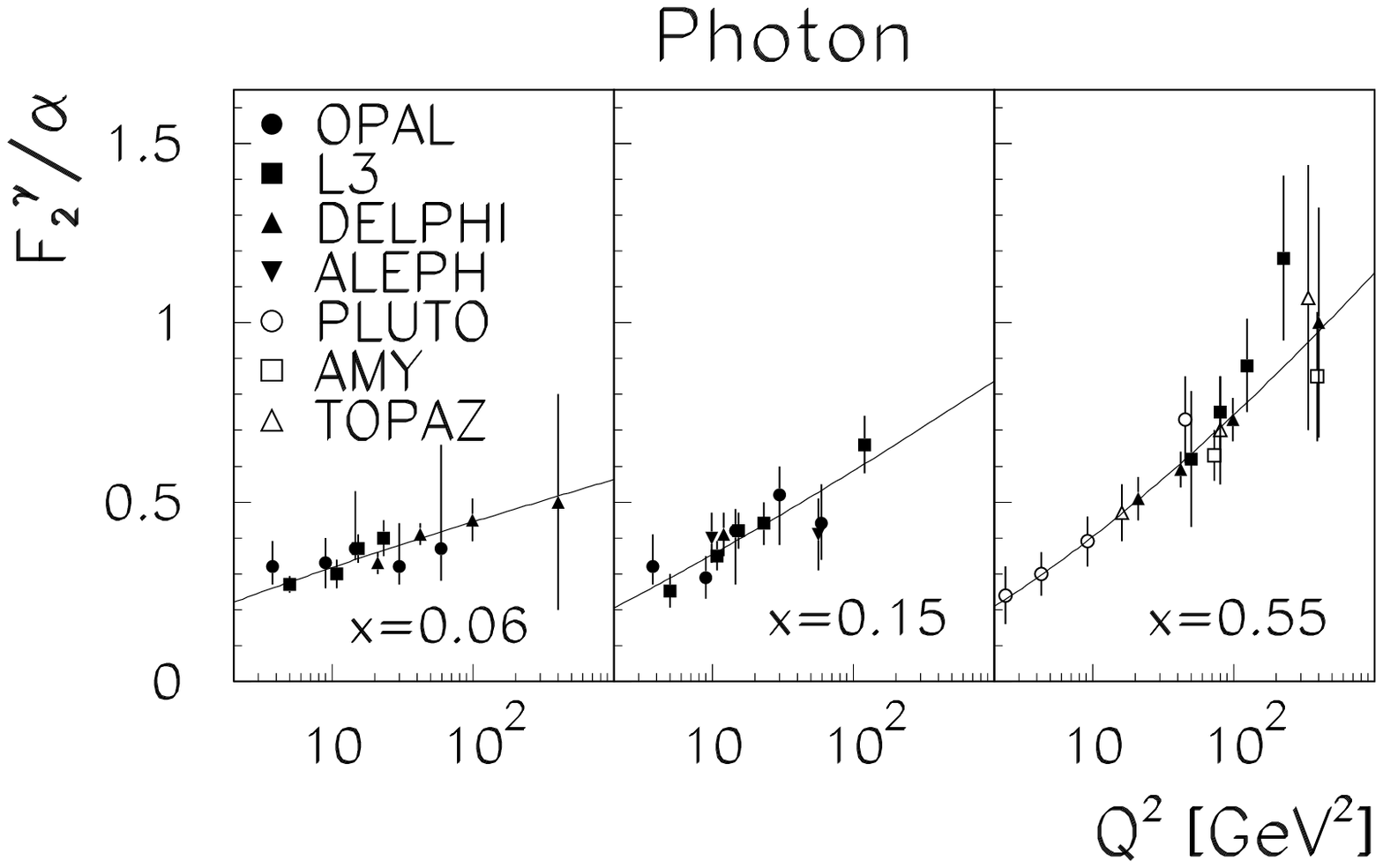, width=15cm}}
\put(1.0,0.0)
{\epsfig{file=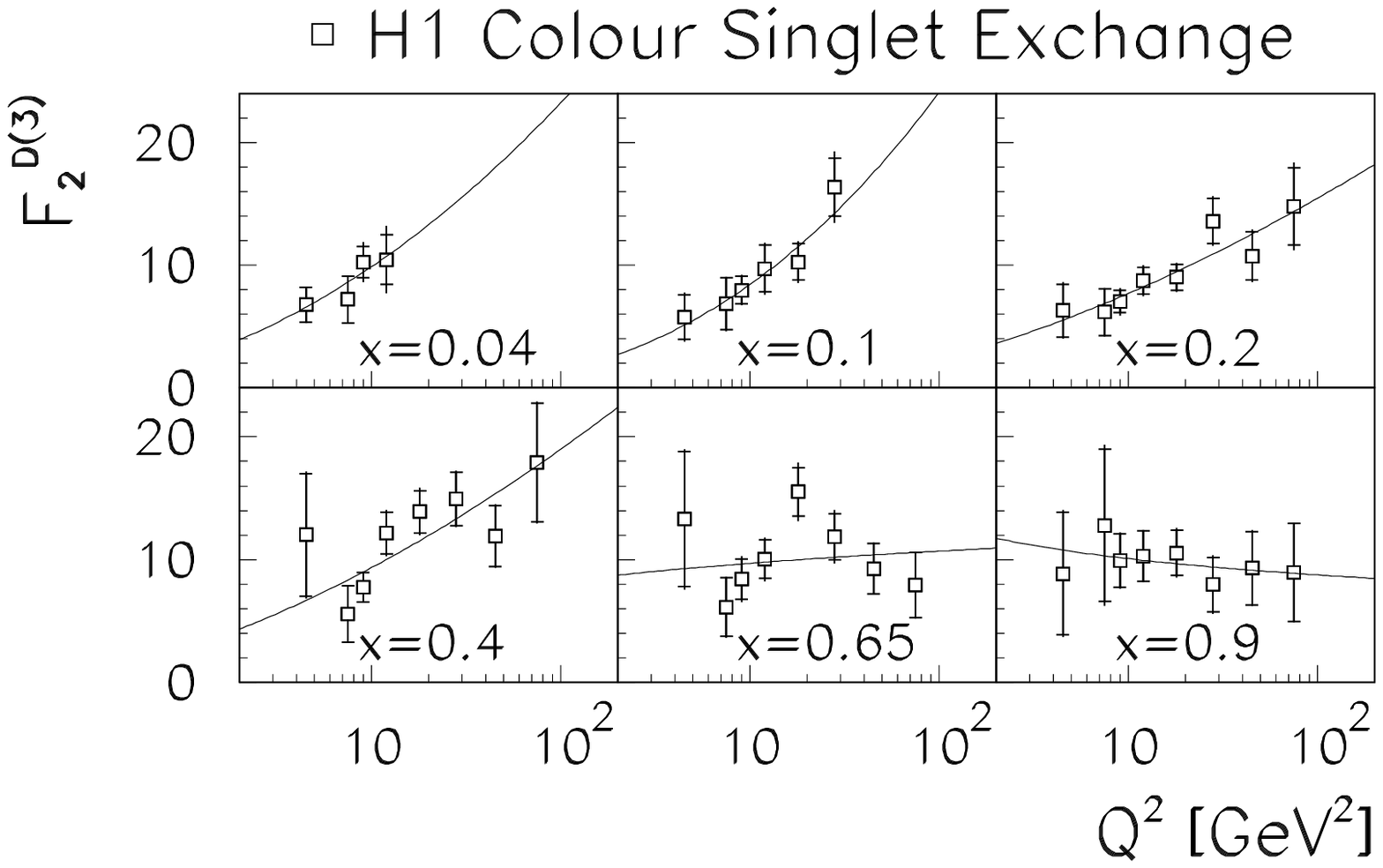, width=15cm}}
\end{picture}
\caption{\label{fig:fluct} 
Measurements of a) the photon structure function $F_2^\gamma/\alpha$ 
and b) the structure function of colour singlet exchange $F_2^{D(3)}$ are
compared to the two-parameter fits of the normalization term $a$ and 
the scaling violation term $\kappa$ according to eq.~(\protect\ref{eq:f2})
using a fixed value of $\Lambda=0.35$ GeV (full curves).}
\end{figure}

For comparison, the same fits have been applied to recent measurements of the
photon structure function $F_2^\gamma$ which have been performed at
$e^+e^-$ colliders \cite{richard} (Fig.\ref{fig:fluct}a, 
$\chi^2/$ndf: $15/37$).
The photon structure results from fluctuations of a photon
into a colour neutral and flavour neutral hadronic state.
The values of the parameters $a$ and $\kappa$ are shown 
in Fig.\ref{fig:akappa} as the open circles.
The photon data exhibit positive scaling violations at all $x$
that are distinct from those of a hadronic bound state like the proton.
The value of $\kappa$ is approximately $1$.
This is as expected from QCD calculations which predict
$F_2^\gamma$ for $0.1 < x < 1$ \cite{witten}.
It is interesting to note that in the low-$x$ region around $x\sim 0.1$ 
the photon data prefer similar values of $a$ to the 
low-$x$ proton data.

In Fig.\ref{fig:fluct}b, H1 structure function measurements $F_2^{D(3)}$ of
colour singlet exchange \cite{diff} are compared to the same two-parameter 
fits as used above ($\chi^2/$ndf: $24/30$).
Here $x$ denotes the fractional momentum of the scattered parton
relative to the colour neutral object, which itself carries a
fractional momentum $\xi=0.003$ relative to the proton.
Also these data exhibit scaling violations $\kappa$ that
are different from the proton measurements at the same values of $x$
(Fig.\ref{fig:akappa}b). 
Instead, at $0.1<x<0.5$ they are large and similar to the photon data
and to the low-$x$ proton data.
The large rate of events with colour singlet exchange together with the
large scaling violations of $F_2^{D(3)}$ is suggestive 
of a gluon dominated exchange.
The values of the normalization $a$ increase with increasing $x$ 
to about $a=10$ as can be seen from extrapolating the curves in
Fig.\ref{fig:fluct}b towards small $Q^2$.
These values have large uncertainties of the order of $100\%$
and are not shown in Fig.\ref{fig:akappa}a.

\section{\boldmath 
Relation of the Parameters with QCD Predictions \label{interpretation}}
\noindent
In the following only the proton data are considered.
The parameter $a$ has already been identified as the 
charge squared weighted quark distributions
extrapolated to $\ln{(Q^2/\Lambda^2)}=1$.
An understanding of the parameters $\Lambda$ and $\kappa$ can be
achieved by comparison with the QCD evolution equation which is 
written here in the leading order DGLAP approximation:
\begin{equation}
\frac{{\rm d} f_i(x,Q^2)}{{\rm d} \ln{Q^2}} =
\frac{\alpha_s(Q^2)}{2\pi} \sum_j \int_x^1 \frac{{\rm d}y}{y} 
P_{ij}\left( \frac{x}{y} \right) f_j(y,Q^2)  \; .
\label{eq:dglap}
\end{equation}
Here $f_i, f_j$ denote the parton densities, $P_{ij}$
are the splitting functions, and
\begin{equation}
\alpha_s = \frac{b}{\ln{(Q^2/\Lambda_{QCD}^2)}}
\label{eq:alphas}
\end{equation}
is the strong coupling constant.

The derivative of the ansatz chosen here, eq.~(\ref{eq:f2}),
with respect to $\ln{Q^2}$ gives
\begin{equation}
\frac{{\rm d} F_2(x,Q^2)}{{\rm d} \ln{Q^2}} 
= \frac{1}{\ln{(Q^2/\Lambda^2)}} \; \kappa(x)  \; F_2(x,Q^2) \; ,
\label{eq:f2t}
\end{equation}
where relating $1/\ln{(Q^2/\Lambda^2)}$ 
with $\alpha_s$ in eq.~(\ref{eq:dglap}) 
implies association of the scale parameter $\Lambda$ in eq.~(\ref{eq:f2}) 
with the QCD parameter $\Lambda_{QCD}$.
The term $\kappa$ relates with the sum over the different 
parton radiation terms in eq.~(\ref{eq:dglap})
divided by $F_2$.
$\kappa$ increases towards small $x$, consistent with larger phase space 
available for parton radiation.

To match the description chosen here, eq.~(\ref{eq:f2}), 
with the double asymptotic approximation expected from QCD for the 
gluon-dominated region at small $x$, 
$F_2\sim \exp{\sqrt{-\ln{x} \; \ln{(\ln{(Q^2/\Lambda^2)})}}}$
\cite{lowx,das},
the scaling violation term $\kappa$ is required to have a dependence like \\
$\kappa\sim \sqrt{-\ln{x}/\ln{(\ln{(Q^2/\Lambda^2)})}}$.
The $Q^2$-dependence of $\kappa$ is therefore expected to be very small
which is in agreement with the experimental observation stated above.

\section{Concluding Remarks}
\noindent
Based on logarithmic $Q^2$-dependence, the presented 
fits of structure function measurements in bins of $x$ give 
information on the hadronic structures and the parton dynamics.

Published proton data, extrapolated to a low resolution scale $Q^2$,
show beyond the valence quarks a sea quark density which is compatible
with a constant value.
Therefore, at small $x$,
the $x$- and $Q^2$-dependencies of $F_2$ dominantly originate from the 
measured scaling violations which are directly related to 
the parton dynamics of the interaction.
More precise data and data reaching smaller values of $x$ will 
determine whether or not the scaling violations further increase 
towards low-$x$ and therefore give valuable information on
the parton densities in the proton as $x$ approaches $0$.

The large scaling violations observed in the structure function data
of colour singlet exchange are suggestive of a gluon dominated exchange.
More precise measurements will test whether at low $Q^2$
one parton carries most of the momentum and 
will provide a reference for a gluon driven regime.

Judgement on a universal low-$x$ behaviour of hadronic structures
will result from measurements of the photon structure function.
If the photon data show a constant quark density at small $x$ 
similar to the low-$x$ proton data, scaling violations of $F_2^\gamma$,
which deviate from those resulting from the photon splitting into 
quark-antiquark pairs and approach those observed 
for the proton, could become visible in the regions around 
$x\sim 0.03$ or $x\sim 10^{-3}$.

\section*{Acknowledgements}

\noindent
For fruitful discussions I wish to thank J.~Dainton, J.~Gayler,
D.~Graudenz,
T.~Naumann, P.~Newman, R.~Nisius, P.~Schleper and F.~Zomer.
I wish to thank Th.~M\"uller and the IEKP group of the University
Karlsruhe for their hospitality, and the Deutsche Forschungsgemeinschaft 
for the Heisenberg Fellowship.

\newpage

\renewcommand{\baselinestretch}{1.0}
{\Large\normalsize}

\end{document}